\documentclass[]{pasj02}

\jyear{2025}
\Received{}
\Accepted{}
 
\usepackage{natbib} 
\usepackage{latexsym}
\usepackage[T1]{fontenc} 
\usepackage[switch,mathlines]{lineno} 
\usepackage{multirow}
\usepackage{mathcomp}
\usepackage{comment}
\usepackage{amsmath}
\usepackage{ulem}
\usepackage{xcolor}%

\begin{document} 

\title{Hadronic origin of the very high-energy gamma-ray emission from the low-luminosity AGN in NGC 4278 }

\author{
 Asahi \textsc{Shoji},\altaffilmark{1}\altemailmark\orcid{0009-0008-8373-2560} \email{asasyo922@gmail.com} 
 Yutaka \textsc{Fujita},\altaffilmark{1}\orcid{0000-0003-0058-9719} 
 Norita \textsc{Kawanaka},\altaffilmark{1}\orcid{0000-0001-8181-7511} 
 Susumu \textsc{Inoue}\altaffilmark{2}\orcid{0000-0003-1096-9424} 
 and 
Kosuke \textsc{Nishiwaki}\altaffilmark{3}\orcid{0000-0003-2370-0475} 
}
\altaffiltext{1}{Department of Physics, Graduate School of Science,
Tokyo Metropolitan University\\
1-1 Minami-Osawa, Hachioji-shi, Tokyo 192-0397}
\altaffiltext{2}{International Center for Hadron Astrophysics, Chiba University\\
1-33 Yayoi-cho, Inage-ku, Chiba City, Chiba 263-8522}
\altaffiltext{3}{INAF Istituto di Radioastronomia, Via P. Gobetti 101, I-40129 Bologna, Italy}




\maketitle

\begin{abstract}

The Large High Altitude Air Shower Observatory has detected very high-energy (VHE) gamma rays from NGC 4278, which is known to host a low-luminosity active galactic nucleus (AGN). Having only very weak radio jets, the origin of its VHE gamma rays is unclear. In this paper we first show that NGC 4278 has a massive molecular cloud surrounding the nucleus by analyzing data taken with the Atacama Large Millimeter/submillimeter Array. We then assume that cosmic ray protons are accelerated in a radiatively inefficient accretion flow around the supermassive black hole, which diffuse into the molecular cloud and produce gamma rays and neutrinos via $pp$ interactions. 
We model the gamma-ray spectra and find that the observations can be explained by such hadronic processes if the AGN activity was higher in the past than at present, and the diffusion coefficient in the molecular cloud is appreciably smaller than in the Milky Way interstellar medium. 
We also show that although the high-energy neutrinos co-produced with the gamma rays are unlikely to be detectable even with IceCube-Gen2, the accompanying synchrotron X-ray emission due to pion-decay secondary electrons and positrons may be detectable in the future, providing a valuable test of our hadronic model.
\end{abstract}


\section{Introduction}

The detection of very high-energy (VHE) gamma rays from active galactic nuclei (AGN) provides crucial insights into the physical processes that occur in these extreme environments. Recently, the Large High Altitude Air Shower Observatory (LHAASO) reported the detection of TeV gamma rays from 1LHAASO J1219+2915 during 2021-2023 \citep{2024ApJS..271...25C,2024ApJ...971L..45C}, which is spatially coincident with NGC 4278, an elliptical galaxy at a distance of $D_{\rm L} = 16.1\: \rm Mpc$ \citep{2001ApJ...546..681T}. 
NGC 4278 harbors a supermassive black hole (SMBH) with a mass of $M_{\rm BH} = (3.09 \pm 0.58)\times 10^8\: M_\odot$ \citep{2003MNRAS.340..793W,2005ApJ...625..716C} and exhibits characteristics of a low-luminosity active galactic nucleus (LLAGN), with a small Eddington ratio\footnote{The mass accretion rate normalized by the Eddington mass accretion rate. See equation (\ref{eq:Ep}).} of $\dot{m} \sim 5\times 10^{-6}$ \citep{2010A&A...517A..33Y,2014A&A...569A..26H}. This discovery is particularly intriguing because the AGN of NGC 4278 is observed to possess only a very weak jet, much less powerful than those for typical gamma-ray sources such as blazars or radio galaxies \citep{2024ApJ...971L..45C}. 
Previous studies have attempted to explain the gamma-ray emission during the active state through various jet-based models, including synchrotron self-Compton (SSC) \citep{2024ApJ...974..134L,2024ApJ...974...56D} and combined SSC + $pp$ interactions \citep{2024ApJS..271...10W}, supported by the detection of two-sided symmetric jets on sub-parsec scales by VLBA observations \citep{2005ApJ...622..178G}. However, crucial properties such as the Doppler factor of the jet is unknown, and it is not obvious how the very weak radio jets can accelerate high-energy particles that can produce the observed VHE gamma rays.

A separate noteworthy fact is that NGC 4278 is known to possess an exceptionally large amount of HI gas on scales of 30-40 kpc, with total mass $\sim 10^8 M_\odot$. This is unusually large for an elliptical galaxy \citep{1981ApJ...246..708R,2006MNRAS.371..157M,2018ApJ...861...49H}.

In this paper, we first show that NGC 4278 has a massive molecular cloud near its center by analyzing data from the Atacama Large Millimeter/submillimeter Array (ALMA). We then propose that the gamma-ray emission from NGC 4278 can be explained by the interaction between cosmic ray (CR) protons\footnote{In this paper, CRs refer to CR protons unless otherwise noted.} accelerated by the AGN and those in the molecular cloud, similar to the model applied to the gamma-ray emission around Sagittarius A$^*$ (Sgr A$^*$; \citealt{2015PhRvD..92b3001F,2017JCAP...04..037F}). 
This mechanism is particularly attractive because NGC 4278, as an LLAGN, is likely to host a radiatively inefficient accretion flow (RIAF). These flows, which form at relatively low mass accretion rates $(\dot{m} \lesssim 0.1\text{--}0.01)$, are characterized by strong turbulence that can potentially accelerate CR protons to PeV energies by stochastic acceleration \citep{2015ApJ...806..159K}. The possible strong turbulence and magnetic fields within the molecular cloud may allow the CR protons to remain trapped and diffuse for a long time, possibly explaining the observed gamma-ray emission.
The gamma-ray emission from NGC 4278 appears to show two distinct states: an active state and a quasi-quiet state, both detected by LHAASO. While Fermi-LAT has detected GeV gamma-ray emission corresponding to the active state \citep{2024ApJ...977L..16B}, only upper limits were obtained during the quasi-quiet state \citep{2022ApJS..260...53A,2023arXiv230712546B,2024ApJ...974..134L}. While the VHE gamma rays appear variable, we focus mainly on the emission during the quasi-quiet state.

We present our modeling of CR diffusion and gamma-ray emission due to $pp$ interactions within the molecular cloud, and our interpretation of the observed gamma rays during the quasi-quiet state. We also explore the detectability of neutrinos with future instruments and emissions due to pion-decay secondary electrons and positrons, which would provide definitive proof of the hadronic origin of the gamma-ray emission.

\section{ALMA observation of NGC~4278}

\subsection{Data Reduction}

NGC~4278 was observed with ALMA at the frequency of the CO($J$=2--1) rotational transition line (230.54 GHz at the rest frame; project code 2022.1.01173.S). The observation was performed with a single pointing, centered on the galaxy on March 9, 2023, with an exposure time of 34 minutes.
The data were calibrated with the appropriate version of the
Common Astronomy Software Application (CASA) software \citep{2007ASPC..376..127M} and the ALMA pipeline for quality assurance.

We subtracted the continuum emission using line-free channels with the CASA task \texttt{uvcontsub} (fitorder = 1) to study the line emissions. We reconstructed the line cubes and the underlying continuum maps with the CASA task \texttt{tclean} with a threshold of $3\:\sigma$. We used natural weighting, following previous studies of elliptical galaxies \citep{2019A&A...631A..22O,2019MNRAS.490.3025R,2023PASJ...75..925F,2024ApJ...964...29F}. The synthesized beam size is $7.1''\times 6.3''$ and the rms in the final data cube is $\sim 9\rm\: mJy\: beam^{-1}$.

\begin{figure*}
  \includegraphics[width=\linewidth]{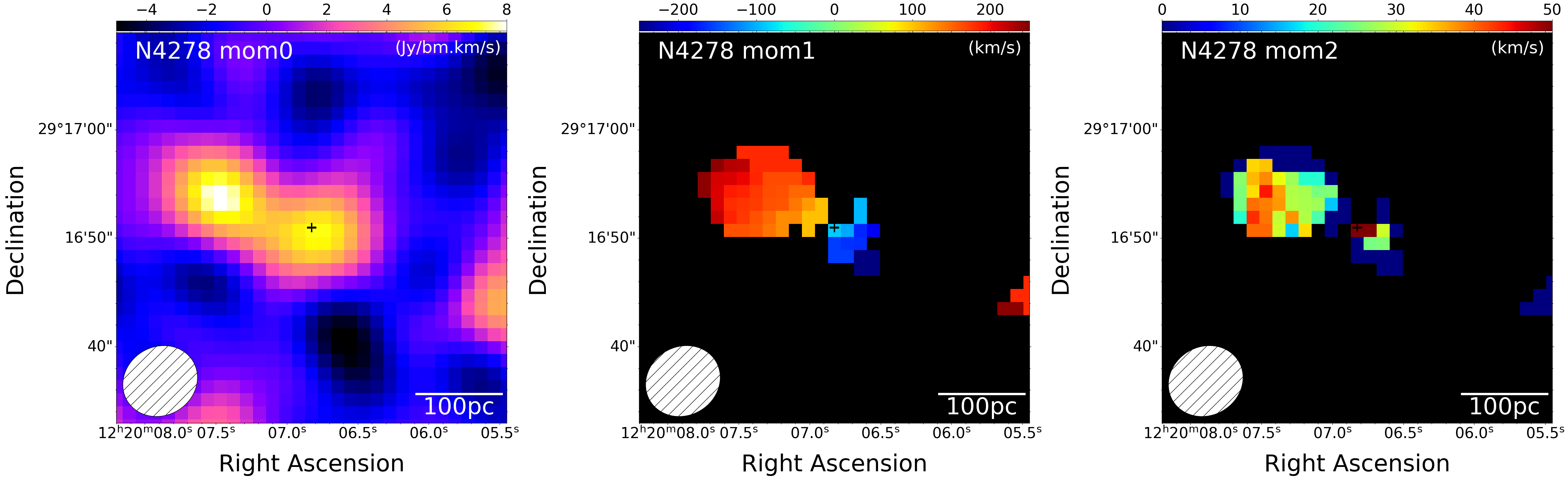} 
\caption{CO emission from NGC 4278. From left to right: integrated intensity map (moment 0), velocity center distribution map (moment 1) of the $\gtrsim 3\sigma$ CO emission, and velocity dispersion map (moment 2) of the $\gtrsim 3\sigma$ CO emission. The position of the AGN is indicated by the cross. The beam size is shown in white at the bottom left of each panel.
{Alt text: Three color maps, each showing the intensity, velocity center and velocity dispersion of the CO emission.}}
\label{fig:ALMA}
\end{figure*}

The CASA task \texttt{immoments} was used to generate images of the integrated intensity, the intensity-weighted mean radial velocity, and the intensity-weighted mean velocity dispersion (full width at half maximum or FWHM). 
The results are shown in figure~\ref{fig:ALMA}. The integrated intensity map in the left panel shows that the CO emission extends to the northeast on a scale of $\sim 150$~pc. The velocity center distribution map in the middle panel suggests rotational motion around the AGN with velocities of $\sim 150$--$200\rm\: km\: s^{-1}$. The axis of rotation is close to the direction of the inner radio jets \citep{2005ApJ...622..178G}.

\subsection{Molecular Gas Mass}

The CO intensity of the elongated cloud around the AGN (figure~\ref{fig:ALMA} left) is $S_{\rm CO}\Delta v \approx 11\:\rm mJy\: beam^{-1}km\: s^{-1}$.
We estimate the mass of the molecular gas from the intensity using the following relationship from \citet{2013ARA&A..51..207B}:
\begin{eqnarray}
 M_{\rm mol} &=& \frac{1.05\times 10^4}{F_{\rm 21}}\left(\frac{X_{\rm CO}}{2\times 10^{20}
\frac{\rm cm^{-2}}{\rm K\: km\: s^{-1}}}\right)
\left(\frac{1}{1+z}\right)\nonumber\\
& &\left(\frac{S_{\rm CO}\Delta v}{\rm Jy\: km\: s^{-1}}\right)
\left(\frac{D_{\rm L}}{\rm Mpc}\right)^2\: M_\odot\:,
\end{eqnarray}
where $X_{\rm CO}$ is the conversion factor from CO to H$_2$ and $D_{\rm L}=16.1$~Mpc is the luminosity distance to NGC~4278. 
The redshift of NGC~4278 is $z\sim 0$.
We take $F_{\rm 21} = 3.2$ for the CO($J$=2--1)/CO($J$=1-0) line flux ratio, as assumed in other studies when this value is not directly measured.
Since the conversion factor $X_{\rm CO}$ is not well understood for elliptical galaxies, we assume a value of $X_{\rm CO}=2\times 10^{20}\rm\: cm^{-2}(K\: km\: s^{-1})^{-1}$ measured in the Milky Way, following previous studies (\citealt{2019A&A...631A..22O,2019MNRAS.490.3025R,2023PASJ...75..925F,2024ApJ...964...29F} and see their discussion). From these values we found $M_{\rm mol}=(9.5\pm 5.5)\times 10^6\: M_\odot$.

\section{Model description}

If protons are accelerated in the AGN at the center of NGC~4278 and subsequently escape, they can interact with the surrounding molecular cloud and emit gamma rays and neutrinos by $pp$ interactions.
We discuss the gamma-ray emission based on a model constructed previously in the context of Sgr A$^*$ by \citet{2015PhRvD..92b3001F} and \citet{2017JCAP...04..037F}.

\subsection{CR acceleration in the RIAF}
\label{ssec:CRaccel}

We assume that CR protons are accelerated in the RIAF. Following the stochastic acceleration model proposed by \citet{2015ApJ...806..159K}, we determine the spectrum of accelerated CRs. The typical energy of the accelerated CRs is calculated by equating their acceleration time in the RIAF to their escape time from the RIAF, which is approximately equal to their diffusion time \citep{2015ApJ...806..159K,2015PhRvD..92b3001F}. The energy is 
\begin{eqnarray}\label{eq:Ep}
    \frac{E_{p,\rm eq}}{m_p c^2} &\sim& 1.4\times 10^{5}
    \left(\frac{\dot m}{0.01}\right)^{1/2}
    \left(\frac{M_{\rm BH}}{1\times 10^{7}M_\odot}\right)^{1/2}
    \left(\frac{\alpha}{0.1}\right)^{1/2}
    \nonumber\\ &\times& \left(\frac{\zeta}{0.1}\right)^{3}
    \left(\frac{\beta}{3}\right)^{-2}
    \left(\frac{R_{\rm acc}}{10 R_S}\right)^{-7/4},
\end{eqnarray}
where $m_p$ is the proton mass, $\dot m = \dot{M}/\dot{M}_{\rm Edd}$ is the gas accretion rate normalized by the Eddington accretion rate, $M_{\rm BH}$ is the mass of the SMBH, $\alpha$ is the alpha parameter for the viscosity in the accretion flow \citep{1973A&A....24..337S}, $\zeta$ is the ratio of the strength of the turbulent fields to that of the non-turbulent fields, $\beta$ is the plasma beta parameter, $R_{\rm acc}$ is the typical radius where CRs are accelerated, and $R_S$ is the Schwarzschild radius of the SMBH.
The Eddington accretion rate is given by $\dot{M}_{\rm Edd}=L_{\rm Edd}/c^2$, where $L_{\rm Edd}$ is the Eddington luminosity.
For NGC 4278 we set the black hole mass to $M_{\rm BH} = 3.1\times 10^{8}\: M_\odot$ \citep{2003MNRAS.340..793W,2005ApJ...625..716C}. In the fiducial model we set $\alpha= 0.1$, $\zeta= 0.05$, $\beta= 3$, and $R_{\rm acc} = 10R_S$, since the diffuse neutrino flux detected at energies of about $10$--$100\: \rm TeV$ by IceCube can be reproduced by the population of LLAGNs with these parameters \citep{2015ApJ...806..159K}. We assume $\dot m = 5\times 10^{-6}$ for the current Eddington ratio of NGC 4278 estimated from X-ray observations \citep{2010A&A...517A..33Y,2014A&A...569A..26H}. With these parameters we obtain $E_{p,\rm eq} = 2.0\: \rm TeV$. Note that we discuss the effects of varying $\dot{m}$, $\zeta$ and $\beta$ later.

The luminosity of CRs accelerated in the RIAF is assumed to be $L_p = \eta_{\rm cr} \dot M c^2$. We choose the parameter $\eta_{\rm cr} = 0.015$ based on \citet{2015ApJ...806..159K}. Considering only stochastic acceleration, the production rate of protons in the momentum range $p$ to $p + dp$ can be expressed as 
\begin{equation}
\label{eq:Nx}
    \dot N (x)dx \propto x^{(7-3q)/2} K_{(b-1)/2}(x^{2-q})dx,
\end{equation}
where $x = p/p_{\rm cut}$, $K_\nu$ is the Bessel function, and $b = 3/(2-q)$, and $q$ represents the spectral index of the turbulent field \citep{2006ApJ...647..539B}. 
We set the lower bound of the CR momentum to $x=m_p c/p_{\rm cut}$. We consider Kolmogorov-type turbulence, so the power-law index is $q = 5/3$. The cutoff momentum can be written as $p_{\rm cut} = (2-q)^{1/(2-q)}p_{\rm eq} = p_{\rm eq}/27$, where $p_{\rm eq} = E_{\rm eq}/c$ \citep{2006ApJ...647..539B,2015ApJ...806..159K}. We normalize equation~(\ref{eq:Nx}) so that the total CR power is $L_p$.

\subsection{Diffusion of CRs in the molecular cloud and gamma-ray emission}
\label{sec:diff}

For simplicity, we assume that the system is spherically symmetric and that CR protons diffuse out from the central AGN into the surrounding molecular cloud. The cloud is uniform and has a radius of $R_{\rm mol}=100$~pc, the approximate apparent size of the actual cloud observed in NGC~4278 (figure~\ref{fig:ALMA} left). Its density is
\begin{align}\label{eq:rho_mol}
  \!\!\rho_{\rm mol}&=\frac{3 M_{\rm mol}}{4\pi R_{\rm mol}^3}\nonumber\\
&=1.5\times 10^{-22}{\rm\: g\: cm^{-3}}\left(\frac{M_{\rm mol}}{9.5\times 10^6\: M_\odot}\right)\left(\frac{R_{\rm mol}}{100\rm\: pc}\right)^{\!\!\!-3}\!\!.
\end{align}
We solve the diffusion equation in spherical symmetry,
\begin{equation}
\label{eq:dfdt}
    \frac{\partial f}{\partial t} = \frac{1}{r^2} \frac{\partial}{\partial r}
    \left(r^2 \kappa \frac{\partial f}{\partial r}\right) + Q \delta^{3}(\boldsymbol r),
\end{equation}
where $f = f(t,r,p)$ is the distribution function of CR protons, $t$ is time, $r$ is the distance from the central AGN, $p$ is the CR momentum, $\kappa$ is the diffusion coefficient, $\delta^{3}(\boldsymbol r)$ is the three-dimensional Dirac delta function, and $Q$ is the CR source term written as $\int 4 \pi c p^3 Q dp = L_p = \eta_{\rm cr} \dot M c^2$. Since the cloud is very large compared to the size of the RIAF, we consider injection by a central point source. We assume that the system is in steady state, so that $\partial f/\partial t =0$. Solving equation~(\ref{eq:dfdt}), we get
\begin{equation}
\label{eq:f}
    f = \frac{Q}{4\pi r \kappa}.
\end{equation}

We assume that the diffusion coefficient of CRs in the molecular cloud is 
\begin{equation}
\label{eq:kappa}
    \kappa(\chi, E_p,B) = 10^{28} \chi\left(\frac{E_p}{10\: \rm GeV}\right)^{1/2}
    \left(\frac{B}{3\: \mu \rm G}\right)^{-1/2}\: \rm cm^{2}\: s^{-1},
\end{equation}
where $\chi$ is the reduction factor, $E_p$ is the CR energy, and $B$ is the magnetic field strength \citep{2009MNRAS.396.1629G}. We assume $B = 1\: \rm mG$, which is consistent with the smaller values observed in molecular clouds near the Galactic Center \citep{2023AJ....166...37G}. Fiducially, we assume $\chi=1$, which would be appropriate for the general interstellar medium (ISM), but not necessarily inside molecular clouds, particularly when a strong CR source is present (see below).

CR protons with sufficient energy can collide with protons in the ambient gas and undergo inelastic $pp$ interactions, leading to the production of neutral and charged pions with roughly equal numbers of $\pi^0$, $\pi^+$ and $\pi^-$. Neutral pions immediately decay into two photons with energy $E_\gamma \simeq 0.1 E_p$, while charged pions eventually decay into neutrinos with energy $E_\nu \simeq 0.05 E_p$, as well as electrons and positrons. The production ratio of electron-flavor neutrinos to muon-flavor neutrinos is 1:2, as long as the muons do not cool significantly before decaying, which is valid for the magnetic field strengths considered here.
We calculate the production rate of gamma rays and neutrinos using the code from \citet{2008ApJ...674..278K} and the parameterization of $pp$ interactions by \citet{2006PhRvD..74c4018K}.
After propagation over extragalactic distances, the ratio among electron, muon and tau neutrinos transitions to 1:1:1 due to neutrino oscillations.

The $pp$ interactions also act as an energy loss mechanism for the protons.
The total inelastic cross section of the $pp$ interaction is well approximated by 
\begin{equation}
\label{eq:sigma}
    \sigma_{pp} = (34.3 + 1.88L + 0.25L^{2}) 
    \left[1 - \left(\frac{E_{\rm th}}{E_p}\right)^{4}\right]^{2}\: \rm mb,
\end{equation}
where $E_{\rm th} = 1.22\: \rm GeV$ is the threshold energy for the production of $\pi^{0}$ mesons and $L = \rm ln$$(E_p/1\: \rm TeV)$ \citep{2006PhRvD..74c4018K}. Thus, the total inelastic cross section at $E_p \sim 1\: \rm TeV$ is $\sigma_{pp} \sim 3.4 \times 10^{-26} \rm cm^2$. Therefore, the $pp$ cooling time of CRs at $E_p \sim 1\: \rm TeV$ is
\begin{eqnarray}
\label{eq:tpp}
    t_{pp} &\sim& (n_{\rm mol} \sigma_{pp} c K_{pp})^{-1}\nonumber\\ &\sim& 6.7 \times 10^{5}\:\left(\frac{n_{\rm mol}}{92\: \rm cm^{-3}}\right)^{-1}\nonumber\\
    &\times& \left(\frac{\sigma_{pp}}{3.4 \times 10^{-26} \rm cm^2}\right)^{-1}
    \left(\frac{K_{pp}}{0.5}\right)^{-1} \rm yr,
\end{eqnarray}
where $n_{\rm mol} = \rho_{\rm mol}/m_p$ is evaluated from equation~(\ref{eq:rho_mol}) and $K_{pp}$ is the proton inelasticity. On the other hand, the diffusion time of CRs in the molecular cloud can be estimated as
\begin{eqnarray}
\label{eq:tdiff}
    t_{\rm diff} &=& \frac{R^{2}_{\rm mol}}{6\kappa}
    \nonumber\\ &\sim& 9.2 \times 10^{4}\: \chi^{-1}
    \left(\frac{R_{\rm mol}}{100\: \rm pc}\right)^2\nonumber\\
    &\times& \left(\frac{E_p}{1\: \rm TeV}\right)^{-1/2}
    \left(\frac{B}{1\: \rm mG}\right)^{1/2} \rm yr,
\end{eqnarray}
using equation~(\ref{eq:kappa}). Since $t_{pp}$ is longer than $t_{\rm diff}$ for $\chi=1$ at the typical energy of the CRs ($E_p \sim 1\: \rm TeV$), we neglect the effect of $pp$ cooling in equations~(\ref{eq:dfdt}) and (\ref{eq:f}) for the fiducial case ($\chi=1$). 

We note that the observed increase in the gamma-ray luminosity of NGC~4278 \citep{2024ApJ...971L..45C} over a period of $\sim 100$~days
cannot be explained by our model because this duration is much shorter than the CR diffusion timescale given in equation~(\ref{eq:tdiff}). The increase may instead be related to the activity of the jet \citep{2024ApJ...971L..45C}. Explanation of the variable gamma-ray component is outside the scope of this paper.

\section{Results}

\subsection{Fiducial model}

\begin{figure}
  \includegraphics[width=8cm]{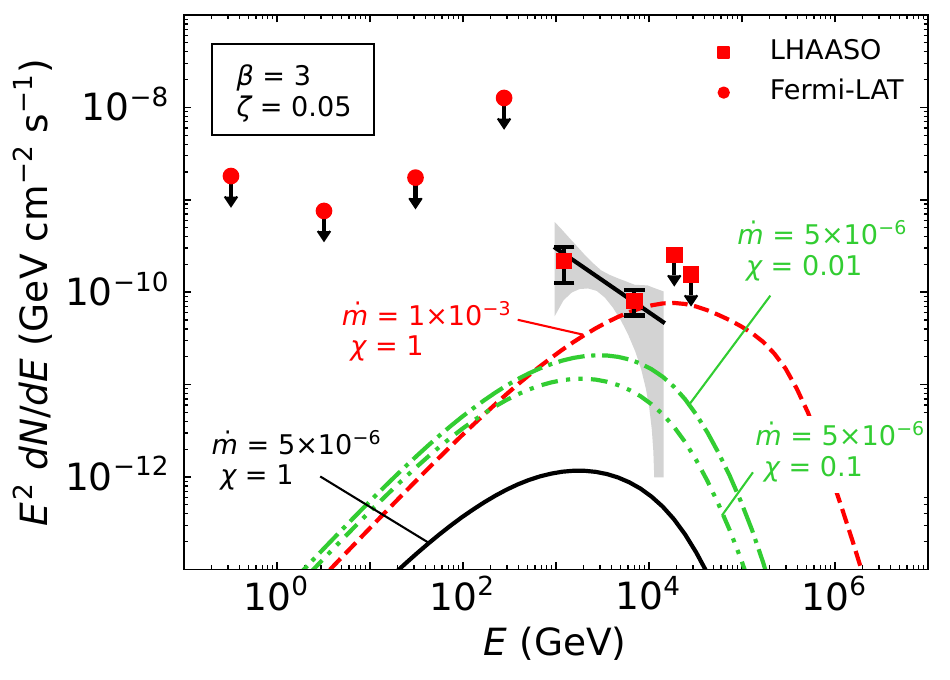} 
\caption{Spectra of the hadronic gamma-ray emission model from the molecular cloud in NGC~4278 for $\dot m = 5\times 10^{-6}$ and $\chi=1$ (black solid), $\dot m = 1\times 10^{-3}$ and $\chi=1$ (red dashed), $\dot m = 5\times 10^{-6}$ and $\chi=0.1$ (green dot-dashed), and $\dot m = 5\times 10^{-6}$ and $\chi=0.01$ (green dot-dot-dashed) when $\beta= 3$ and $\zeta= 0.05$. Filled squares and circles show the data from LHAASO  \citep{2024ApJ...971L..45C} and Fermi \citep{2024ApJS..271...10W}, respectively.
{Alt text: Line graph. The x axis shows the photon energy from $10^{-1}$ to $10^{7}$ giga electron volts. The y axis shows the energy flux from $10^{-13}$ to $10^{-7}$ giga electron volts per squared centimeter per second.}}
\label{fig:mol_1}
\end{figure}

Figure~\ref{fig:mol_1} shows the model spectra of gamma-ray emission from the molecular cloud for the fiducial case ($\dot m = 5\times 10^{-6}$, $\chi=1$, $\zeta= 0.05$, and $\beta= 3$), compared with the observed data from LHAASO \citep{2024ApJ...971L..45C} and Fermi \citep{2024ApJS..271...10W}. 
Since gamma-ray attenuation due to pair production with the extragalactic background light (EBL) can be effective for VHE gamma rays, the effect is included in the LHAASO data using the EBL model of \citet{2021MNRAS.507.5144S}. Figure~\ref{fig:mol_1} shows that the fiducial model (black solid curve) significantly falls short of the observations.

\subsection{Enhanced past activity and CR confinement}
Considering the simplicity of our model and the large uncertainties expected in some of the parameters, we vary them from the fiducial values to see if the model can be made consistent with the observations.

First, we explore the possibility of a larger $\dot m$. The AGN at the center of the Milky Way, Sgr A$^*$, is also a LLAGN. Its current mass accretion rate is $\dot m = 4.2\times 10^{-6}$ \citep{2015PhRvD..92b3001F}, similar to that of NGC 4278 ($\dot{m} \sim 5\times 10^{-6}$; \cite{2010A&A...517A..33Y,2014A&A...569A..26H}).
However, previous studies suggest that $\dot m$ of Sgr A$^*$  was $10^{3}$--$10^{4}$ times larger at epochs more than 100 years ago \citep{1996PASJ...48..249K,2006PASJ...58..965T,2013PASJ...65...33R}. 
\citet{2015PhRvD..92b3001F} showed that the gamma-ray emission observed from the molecular clouds around Sgr A* can be explained if the increased CR production associated with the past enhanced activity is taken into account.

For NGC 4278, X-ray observations over the last 20 years reveal periods when the X-ray flux was about an order of magnitude higher, plausibly due to a correspondingly higher $\dot m$, compared to later epochs when $\dot{m} \sim 5\times 10^{-6}$ was inferred \citep{2024ApJ...974..134L}. It is possible that $\dot m$ was even higher on longer timescales in the past, which may possibly be related to the presence of massive molecular and HI gases.
Thus, we consider the case where the AGN activity and CR production of NGC~4278 was much higher in the past and has recently become more quiet.
As long as the quiescent period is shorter than the CR diffusion time, $t_{\rm diff} \sim 10^{5}\: \rm yr$ (equation~(\ref{eq:tdiff})), the current gamma-ray flux from the molecular cloud is relatively insensitive to the recent decrease in CR production. 

The gamma-ray emission for the case of $\dot m = 1\times 10^{-3}$ with other parameters remaining fiducial is shown by the red dashed curve in figure~\ref{fig:mol_1}. Compared to the fiducial case (black solid curve), the flux around $E \approx 10$ TeV is increased significantly.
However, according to equation~(\ref{eq:Ep}), larger $\dot m$ results in larger typical CR energy $E_{p,\rm eq}$, and the peak of the gamma-ray spectrum is shifted toward higher energies; for $\dot m = 1\times 10^{-3}$, $E_{p,\rm eq} = 29\: \rm TeV$ (equation~(\ref{eq:Ep})). 
This model is not compatible with the observed LHAASO spectrum at $E = 1- 10$ TeV.

Second, we consider the possibility of a smaller diffusion coefficient inside the molecular cloud, since the nature of the relevant magnetic turbulence that govern CR propagation is uncertain and in principle can be very different from the general ISM, especially near a strong CR source. Various observations indicate that the diffusion coefficient in the vicinity of potential CR sources could be one to two orders of magnitude smaller than the value inferred for the Milky Way ISM, such as in the Cygnus bubble \citep{2024SciBu..69..449L,2024ApJ...974..276N}, around the microquasar SS 433 \citep{2024arXiv241008988L}, and in the pulsar TeV halos \citep{2017Sci...358..911A,2022ApJ...936..183B}. The magnetohydrodynamical waves self-generated by the CRs escaping from the source could be the origin of the small diffusion coefficient \citep{2009ApJ...707L.179F,2010ApJ...712L.153F,2022MNRAS.512..233S}.
Therefore, we reduce the coefficient from $\chi = 1$ in equation~(\ref{eq:kappa}), keeping the other parameters fiducial. Note that the dependence on $\chi$ and $B$ is degenerate in equation~(\ref{eq:kappa}), and the results are identical for given $\chi B^{-1/2}$.
A smaller $\chi$ implies a longer $t_{\rm diff}$ (equation~(\ref{eq:tdiff})), which could exceed $t_{pp}$ (equation~(\ref{eq:tpp})). 
For simplicity, we include the effect of $pp$ cooling by ignoring the CRs at $r>R_{\rm cool}$, where $R_{\rm cool}$ is the cooling radius at which $t_{pp}=t_{\rm diff}$.
Specifically, equation~(\ref{eq:f}) is changed to
\begin{equation}\label{eq:fcool}
    f =
    \begin{cases}
     \frac{Q}{4\pi r \kappa(\chi, E_p,B)} & \text{if $r < R_{\rm cool}$,} \\
     0                       & \text{if $r \geq  R_{\rm cool}$,}
    \end{cases}
\end{equation}
\begin{equation}\label{eq:Rcool}
    R_{\rm cool} = \sqrt{6 \kappa t_{pp}}\:.
\end{equation}
Figure~\ref{fig:mol_1} shows the gamma-ray spectra for $\chi = 1$ (black solid curve), 0.1 (green dot-dashed curve), and $0.01$ (green dot-dot-dashed curve). Comparing the cases of $\chi = 1$ and 0.1, a smaller $\chi$ results in a larger gamma-ray flux for this range of $\chi$. Since we assume that $\rho_{\rm mol}$ is constant at $r<R_{\rm mol}$, the gamma-ray flux at a given energy $E$ saturates when $R_{\rm cool} < R_{\rm mol}$ for the CRs with $E_p \approx 10E$.
Note that $R_{\rm cool}$ increases with $E_p$ due to the energy dependence of $\kappa$ (equations~(\ref{eq:kappa}) and (\ref{eq:Rcool})).
For our model with $\dot m = 5\times 10^{-6}$, $\beta = 3$, $\zeta = 0.05$, the gamma-ray flux around $E = 1-10$ TeV saturates at $\chi \sim 0.01$. The maximal flux is about an order of magnitude lower than the observed values at $E\sim 10$~TeV. Thus, the reduced $\chi$ alone cannot explain the data.

\begin{figure}
  \includegraphics[width=8cm]{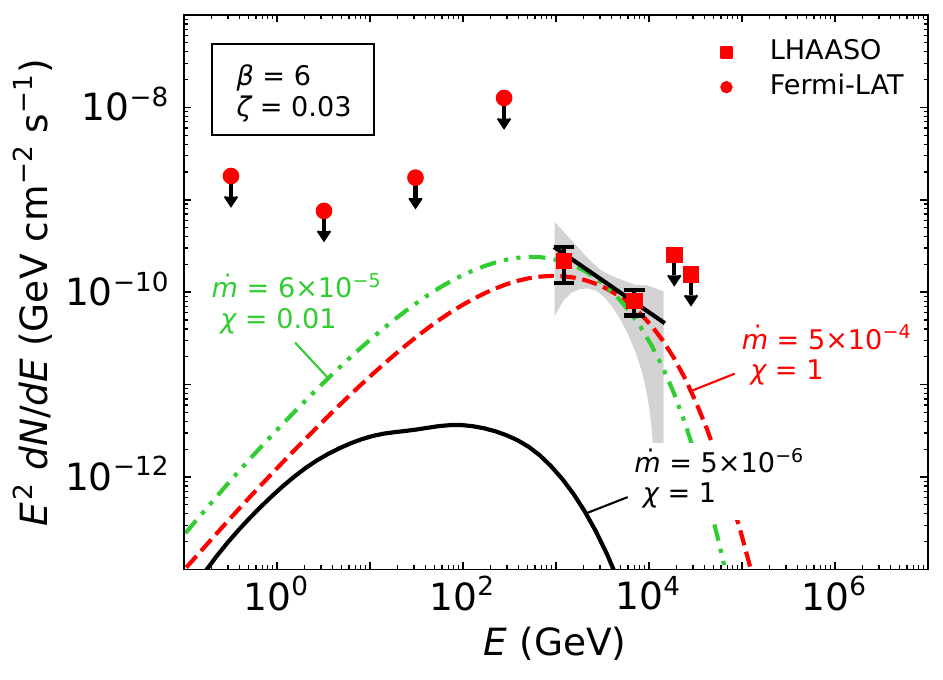} 
\caption{Model gamma-ray spectra for $\dot m = 5\times 10^{-6}$ and $\chi=1$ (black solid), $\dot m = 5\times 10^{-4}$ and $\chi=1$ (red dashed), and $\dot m = 6\times 10^{-5}$ and $\chi=0.01$ (green dot-dashed) when $\beta = 6,\: \zeta = 0.03$. Otherwise similar to figure~\ref{fig:mol_1}.
{Alt text: Line graph similar to figure~\ref{fig:mol_1}}.}
\label{fig:mol_2}
\end{figure}

Third, we modify the parameters related to the CR acceleration model explained in section~\ref{ssec:CRaccel}. 
We focus on the plasma beta parameter $\beta$ and the turbulence strength parameter $\zeta$, as these are very uncertain theoretically as well as observationally, but can significantly affect the CR spectrum (equation~(\ref{eq:Ep})).
Figure~\ref{fig:mol_2} shows the model gamma-ray flux when we vary $\beta$ from 3 to 6, and $\zeta$ from 0.05 to 0.03, so that $E_{p,\rm eq}$ is less than the fiducial value by a factor of 20.
Note that $E_{p,\rm eq} \propto \zeta^3\beta^{-2}$ (equation~(\ref{eq:Ep})), so the dependence on $\zeta$ and $\beta$ is degenerate. The typical CR energy is $E_{p,\rm eq} = 0.11\: \rm TeV$ for $\dot m = 5\times 10^{-6}$, $E_{p,\rm eq} = 0.38\: \rm TeV$ for $\dot m = 6\times 10^{-5}$, and $E_{p,\rm eq} = 1.1\: \rm TeV$ for $\dot m = 5\times 10^{-4}$ (equation~(\ref{eq:Ep})). The observed gamma-ray flux at $E \sim 1$--$10\: \rm TeV$ is well reproduced when $\dot{m} = 5\times 10^{-4}$ and $\chi = 1$ (red dashed curve), or $\dot{m} = 6\times 10^{-5}$ and $\chi = 0.01$ (green dot-dashed curve). This suggests the existence of an allowed range of $\dot{m}$ and $\chi$ for given $\beta$ and $\zeta$ which determine $E_{p,\rm eq}$ or the peak energy of the gamma-ray spectrum.
Figure~\ref{fig:parameter} shows the combination of $\dot m$ and $\chi$ that can reproduce the observed spectrum when $\beta=6$ and $\zeta=0.03$. The range of $6 \times 10^{-5} \lesssim \dot m \lesssim 7 \times 10^{-4}$ is consistent with gamma-ray observations when $\chi<1$ (dashed line). The value of $\dot{m}$ indicates that the past activity of NGC 4278 may have been 10 to 100 times greater than at present ($\dot{m}=5\times 10^{-6}$).

\subsection{HI gas}
NGC 4278 is known to have an exceptionally large amount of HI gas for an elliptical galaxy, with total mass $\sim 10^8 M_\odot$ \citep{1981ApJ...246..708R,2006MNRAS.371..157M,2018ApJ...861...49H}. It appears as a spheroid with major and minor axes of $\sim 37$~kpc and $\sim 30$ kpc, respectively \citep{2006MNRAS.371..157M}. We calculate the contribution of the HI gas to the gamma-ray flux.

We set the HI gas mass to $M_{\rm HI} = 3.5 \times 10^8 M_\odot$ \citep{2018ApJ...861...49H}. We assume that it takes the form of a sphere with radius $R_{\rm HI} \sim 15$~kpc. The mean density is
\begin{align}\label{eq:rho_HI}
  \!\rho_{\rm HI}&=\frac{3 M_{\rm mol}}{4\pi R_{\rm HI}^3}\nonumber\\
&=1.7\times 10^{-27}{\rm\: g\: cm^{-3}}\left(\frac{M_{\rm HI}}{3.5\times 10^8\: M_\odot}\right)\!\left(\frac{R_{\rm HI}}{15\rm\: kpc}\right)^{\!\!\!-3}\!\!\!.
\end{align}

The effect of $pp$ cooling on CRs in the HI gas can be estimated by equations~(\ref{eq:fcool}) and (\ref{eq:Rcool}) with $\kappa$ (equation~(\ref{eq:kappa})) and $t_{pp}$ (equation~(\ref{eq:tpp})) using parameters appropriate for the HI gas. We assume $B = 50\: \rm \mu G$ for the HI gas, which is similar to the values observed for intercloud regions around the Galactic Center of the Milky Way \citep{2011IAUS..271..170F}.

\begin{figure}
  \includegraphics[width=8cm]{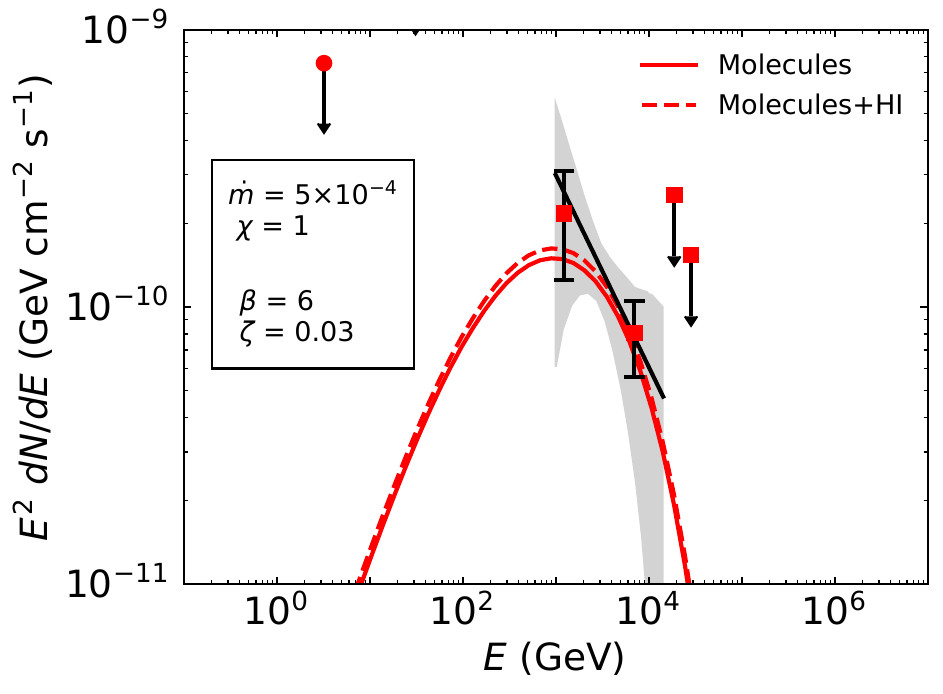} 
\caption{Model gamma-ray spectra from the molecular cloud only (solid) and from both the molecular cloud and the HI gas (dashed) when $\dot m = 5\times 10^{-4}$ and $\chi = 1$. Otherwise similar to figure~\ref{fig:mol_1}.
{Alt text: Line graph similar to figure~\ref{fig:mol_1}}.}
\label{fig:HI}
\end{figure}

\begin{figure}
  \includegraphics[width=8cm]{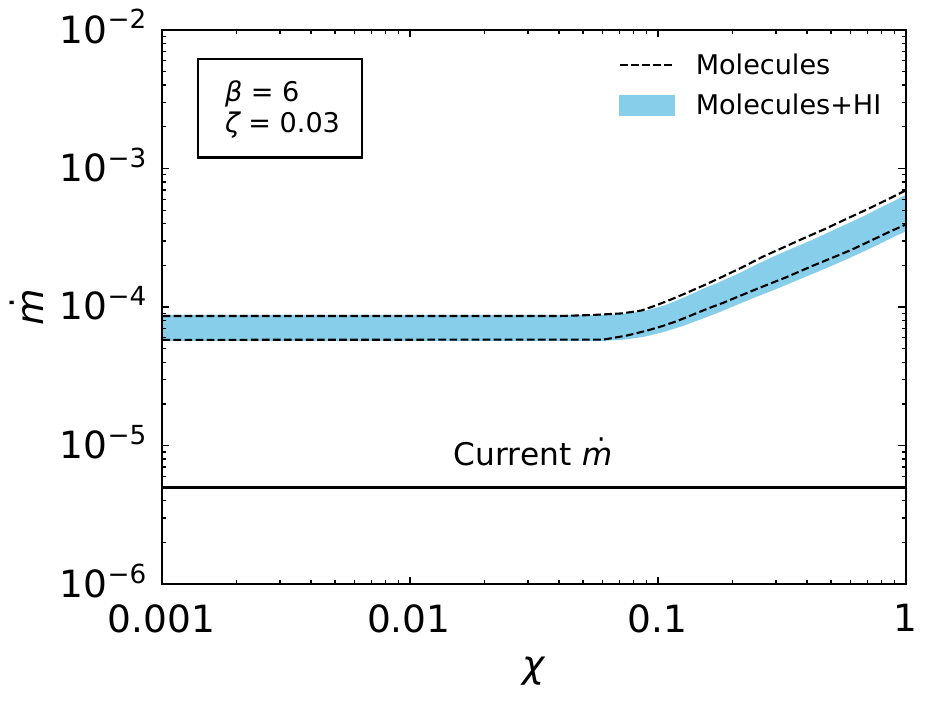} 
\caption{The combination of $\dot m$ and $\chi$ that can reproduce the observed gamma-ray data for the molecular cloud only (dashed) and for both the molecular cloud and the HI gas (cyan shaded region). The horizontal line represents the current accretion rate, $\dot m = 5 \times 10^{-6}$. 
{Alt text: Graph of the parameter space. The x axis shows $\chi$ from $0.001$ to $1$. The y axis shows $\dot m$ from $10^{-6}$ to $10^{-2}$.}}
\label{fig:parameter}
\end{figure}

Figure~\ref{fig:HI} shows the result when $\dot m = 5\times 10^{-4}\:, \chi = 1\:, \beta = 6,\:$ and $\zeta = 0.03$. The contribution of the HI gas is less than $1/10$ of that of the molecular cloud. Figure~\ref{fig:parameter} shows the allowed combination of $\dot m$ and $\chi$. The range of $6 \times 10^{-5} \lesssim \dot m \lesssim 6 \times 10^{-4}$ is consistent with gamma-ray observations.

\subsection{Neutrino emission}
Neutrinos are inevitably co-produced with gamma rays in $pp$ interactions, with a spectrum roughly mirroring that of the latter. The detection of high-energy neutrinos would strongly support our hadronic model. We calculate the neutrino flux employing the same code \citep{2008ApJ...674..278K} and parameterization \citep{2006PhRvD..74c4018K} as for the gamma rays.

\begin{figure}
  \includegraphics[width=8cm]{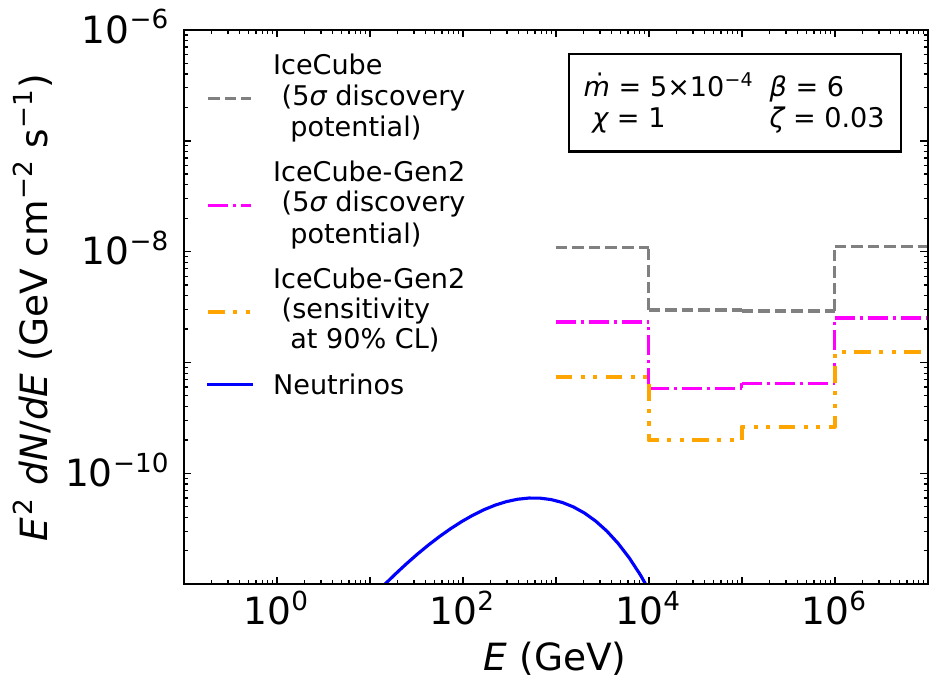} 
\caption{The muon neutrino flux from the molecular cloud and the HI gas when $\dot m = 5\times 10^{-4}$ and $\chi = 1$. The differential sensitivities for the detection of point sources (5$\sigma$ discovery potential and sensitivity at 90\% CL for track-like events) of IceCube and IceCube-Gen2 are also shown for the declination $\delta = 30\tcdegree$ \citep{2021JPhG...48f0501A}.
{Alt text: Line graph.}}
\label{fig:neutrino}
\end{figure}

In figure~\ref{fig:neutrino}, the blue curve shows the predicted {muon} neutrino flux when $\dot m = 5\times 10^{-4}$, $\chi = 1,\: \beta = 6$, and $\zeta = 0.03$, corresponding to the gamma-ray emission model shown as the dashed curve in figure~\ref{fig:HI}. The figure also shows the detectability of neutrinos for the declination of NGC 4278 with IceCube and IceCube-Gen2, in terms of the differential point source sensitivity at 90\% CL and 5$\sigma$ discovery potential {for track-like events} \citep{2021JPhG...48f0501A}. The sensitivity at 90\% confidence level (CL) is defined as the neutrino flux required for the test statistic (TS) distribution to exceed the median of the background-only TS distribution 90\% of the time. The 5$\sigma$ discovery potential is defined as the neutrino flux required for the TS to exceed the 5$\sigma$ fluctuation of the background-only TS distribution 50\% of the time \citep{2024ApJ...976....8A}. The assumed declination is $\delta = 30\tcdegree$, close to that of NGC 4278 ($\delta \sim 29.28\tcdegree$; \cite{2019ApJS..242....5X}). The predicted neutrino flux is much lower than the sensitivity of IceCube-Gen2. Therefore, neutrino detection will be difficult in the near future.

\section{Discussion}

\subsection{Secondary radiation}

The decay of charged pions generate secondary electrons and positrons. Synchrotron and inverse Compton radiation from these secondary particles may contribute to the broadband emission. Radio to UV emission has been observed from regions within roughly 10 pc around the BH of NGC 4278 \citep{2005ApJ...622..178G,2009A&A...508..641C}, which may suggest that such emission originates from the AGN rather than the molecular cloud that has an extension of $\sim 100\: \rm pc$. On the other hand, the emission from secondary particles produced in the molecular cloud may contribute to the observed X-ray emission, whose origin is only constrained to be within 200 pc from the AGN \citep{2010A&A...517A..33Y,2012ApJ...758...94P}.

We calculate the expected broadband emission due to secondary particles using the same code as for the gamma rays and neutrinos. We assume that the cosmic microwave background dominates the seed photons for inverse-Compton scattering. The interstellar radiation field is likely to be less important, as the star formation rate of NGC 4278 is appreciably lower than that of the Milky Way \citep{2017MNRAS.464..329Y}.

\begin{figure}
  \includegraphics[width=8cm]{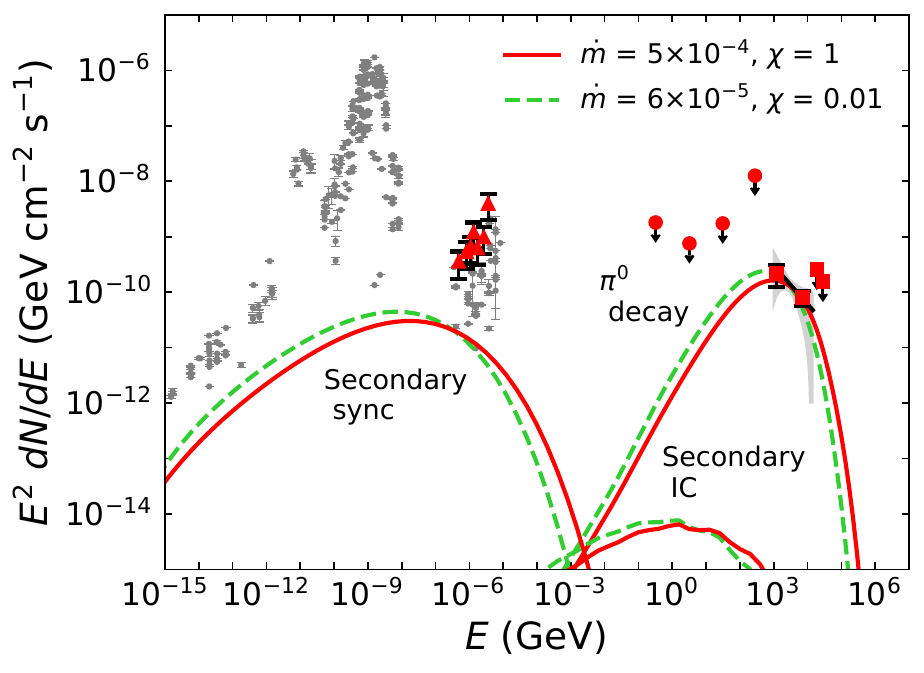} 
\caption{The broadband emission from neutral pion decay, secondary synchrotron, and secondary inverse Compton for $\dot m = 5\times 10^{-4}$ and $\chi = 1$ (red solid) and $\dot m = 6\times 10^{-5}$ and $\chi = 0.01$ (green dashed) when $\beta = 6$ and $\zeta = 0.03$. Red squares, circles, and triangles show the data from LHAASO \citep{2024ApJ...971L..45C}, Fermi-LAT, and Swift-XRT \citep{2024ApJS..271...10W}, respectively. Gray circles show archival data from NASA/IPAC Extragalactic Database catalogs.
{Alt text: Line graph.}}
\label{fig:secondary}
\end{figure}

Figure~\ref{fig:secondary} shows the secondary synchrotron and inverse-Compton emission for $\dot m = 5\times 10^{-4}$ and $\chi = 1$ (red solid curve), and $\dot m = 6\times 10^{-5}$ and $\chi = 0.01$ (green dashed curve) when $\beta = 6$ and $\zeta = 0.03$. For these model parameters, secondary synchrotron emission is seen to span the radio to X-ray bands, but is generally subdominant compared to the observed data. Within the range of $\dot m$ and $\chi$ shown in Figure~\ref{fig:parameter} that can reproduce the gamma-ray observations, the spectral shape of the secondary emission remains almost unchanged. Secondary inverse Compton emission reaches the gamma-ray band, but is completely negligible relative to the $\pi^0$ decay gamma rays. Nevertheless, depending on the parameters, the secondary contributions can become observationally interesting, especially in X-rays.
Since the emission from the RIAF or jet can vary on timescales of years or less, observations during periods when such emission is sufficiently faint may reveal the secondary radiation from the molecular cloud, providing crucial supporting evidence for the hadronic origin of the gamma rays.
Furthermore, it could offer unique and valuable constraints on physical quantities such as the magnetic fields in the molecular cloud.

\subsection{Other LLAGNs}

In addition to NGC 4278, LHAASO has detected two other objects that are sometimes classified as LLAGNs; NGC 1275 \citep{2025MNRAS.540.1860C} and M87 \citep{2024ApJ...975L..44C}. Only flares with clear variability have been observed from NGC 1275, indicating a gamma-ray emission site on subparsec scales \citep{2025MNRAS.540.1860C}. M87 has been observed during both flaring and quiescent states, but it possesses almost no molecular gas \citep{2025ApJ...990...66K}. Thus, our model may not be relevant for the cases of NGC 1275 and M87.

Besides the recent finding of GeV emission from NGC 4278, four LLAGN have been detected by Fermi-LAT: NGC 315, NGC 1275, NGC 4261, and M87 \citep{2020MNRAS.492.4120D}. With the exception of M87, molecular clouds with masses of $\sim 10^7$--$10^8 M_\odot$ have been observed in these LLAGNs \citep{2021ApJ...908...19B,2019ApJ...883..193N}. This motivates us to investigate the origin of the gamma-ray emission from these objects in the context of our model in the future.

\section{Conclusions}

We have investigated the origin of the very high-energy gamma rays from NGC 4278 detected by LHAASO. First, we analyzed ALMA data and found that the galaxy has a massive ($\sim 10^7\: M_\odot$) molecular cloud within a few 100 pc from its center. Then we constructed a theoretical model in which CR protons are stochastically accelerated in a RIAF around the central SMBH with power proportional to the accretion rate. The CRs escaping from the RIAF enter the surrounding molecular cloud and produce gamma rays and neutrinos through $pp$ interactions.

We calculated the gamma-ray flux from the molecular cloud and found that it falls short of that observed by LHAASO if the accretion rate onto the SMBH has persisted at the currently estimated rate over timescales of $10^5\: \rm yr$ relevant for CR diffusion.
Therefore, we considered models with an enhanced past accretion rate and a suppressed diffusion coefficient for CR transport. The former is motivated by observations of Sgr A$^*$, an LLAGN known to have had significantly higher past activity, as well as past X-ray observations of NGC 4278. The latter can be realized if the turbulence inside the molecular cloud is stronger than in the general ISM, possibly generated by the escaping CRs themselves. 
We showed that the observed gamma-ray flux can be reproduced if the accretion rate was $\sim 10$--100 times higher in the past and the diffusion coefficient is $\lesssim 0.1$--1 times smaller than the value inferred for the ISM.

We also calculated the gamma rays from the HI gas outside the molecular cloud of NGC 4278 and found that its contribution are much smaller than that of the molecular cloud.
We also found that the $pp$ neutrinos co-produced with the gamma rays from the molecular cloud and the HI gas of NGC 4278 are unlikely to be observable even with the future neutrino observatory IceCube-Gen2. 
Finally, the emission due to secondary electrons and positrons generated in $pp$ interactions was found to be generally subdominant compared with the available data. However, future observations during epochs of low AGN activity may reveal the secondary synchrotron X-rays, which would support the hadronic origin of the gamma rays and probe the magnetic fields in the molecular cloud.

\begin{ack}
This work was supported by NAOJ ALMA Scientific Research grant Code 2022-21A, and JSPS KAKENHI Grant Nos. JP22H01268, JP22H00158, JP22K03624, JP23H04899 (Y.F.); JP22K03686, JP25K01045 (N.K.); JP23H04896 (S.I).
This paper makes use of the ALMA data: ADS/JAO.ALMA\#2022.1.01173.S.
ALMA is a partnership of ESO (representing its member states), NSF (USA) and NINS (Japan), together with NRC (Canada), MOST and ASIAA (Taiwan), and KASI (Republic of Korea), in cooperation with the Republic of Chile.
The JAO is operated by ESO, AUI/NRAO and NAOJ.
Data analysis was in part carried out on the Multi-wavelength Data Analysis System operated by the Astronomy Data Center (ADC), National Astronomical Observatory of Japan.
\end{ack}

\end{document}